# On the anomalous density for Bose gases at finite temperature

A. Boudjemâa[1] and M. Benarous

*Laboratory for Theoretical Physics and Material Physics*
*Faculty of Sciences and Engineering Sciences*
*Hassiba Benbouali University of Chlef*
*B.P. 151, 02000, Chlef, Algeria.*

**Abstract**

We analyze the behavior of the anomalous density as function of the radial distance at different temperatures in a variational framework. We show that the temperature dependence of the anomalous density agrees with the Hartree-Fock-Bogoliubov (HFB) calculations. Comparisons between the normal and the anomalous fractions at low temperature show that the latter remains higher and consequently the neglect of the anomalous density may destabilize the condensate. These results are compatible with those of Yukalov. Surprisingly, the study of the anomalous density in terms of the interaction parameter shows that the dip in the central density is destroyed for sufficiently weak interactions. We explain this effect.

---

[1] Corresponding author :e-mail: **a.boudjemaa@univ-chlef.dz**

# On the anomalous density for Bose gases at finite temperature

1. **Introduction**

The investigation of Bose-Einstein condensation phenomenon at finite temperature has attracted a great number of physicists in the recent years. The traditional theories to describe the Bose-Einstein condensation at finite temperature are based on the Bogoliubov quasiparticle approach developed originally for a spatially homogeneous Bose-condensed gas at $T \to 0$ [1] and employed by Lee and Yang [2] at finite temperatures. The generalization of the Bogoliubov method for spatially inhomogeneous systems has been described by De Gennes [3]. The main idea of Bogoliubov is to separate out the condensate part in the field operator $\Psi(\vec{r}) = \Phi(\vec{r}) + \overline{\Psi}(\vec{r})$. The system may be split into two sub-components, namely the condensate described by its density $n_c(r) = |\Phi(r)|^2$ and the thermal cloud $\tilde{n}(\vec{r}, \vec{r}') = <\overline{\Psi}^+(\vec{r})\overline{\Psi}(\vec{r}')>$. These two components have been intensively studied both theoretically and experimentally [4-34]. Similarly, this approximation motivates the definition of an additional mean field contribution $\tilde{m}(\vec{r}, \vec{r}') = <\overline{\Psi}(\vec{r})\overline{\Psi}(\vec{r}')>$. This is often referred to as the pair anomalous average, and bears its name from the fact that there is an unequal number of creation and annihilation operators being averaged over. An analogous correlation plays a dominant role in the BCS theory of superconductivity [3], where fermionic atoms pair up to form the so-called Cooper pairs. In the case of Bose-Einstein condensation of neutral bosonic atoms, where the condensate mean field $\Phi(r)$ is the dominant parameter, the anomalous average plays a minor role for sufficiently repulsive interactions atoms in the condensate. This contribution becomes however crucial in the presence of attractive interactions and molecular BECs [35].

Among the theoretical investigations of the anomalous density, we can cite in particular those of Hutchinson et al [36] and Giorgini [37] based on the mean field HFB-BdG approximation. To go beyond the mean field, Fedichev et al [38] and Proukakis et al [39, 40] developed a finite temperature perturbation theory using an HFB basis. Another kind of approaches has been developed by Griffin [41] based on the Green's function method to derive an equation for the condensate and its fluctuations. Yukalov [42]

# On the anomalous density for Bose gases at finite temperature

adopted a quite different approach by using the notion of representative statistical ensembles. Recently, Wright et al. [43, 44] have used the classical-field trajectories of the Projected Gross-Pitaevskii equation.

An interesting alternative non-perturbative and non-classicalfield approach to the finite temperature Bose gas is provided by the so-called time-dependent variational principle. This principle was proposed by Balian and Vénéroni (BV) a long time ago [45]. The BV variational principle has been applied to various quantum problems including heavy ion reactions [46], quantum fields out of equilibrium [47] attempts to go beyond the Gaussian approximation for fermion systems [48]. Therefore, the BV principle has been used to provide the best approximation to the generating functional for two and multi-time correlation functions of a set of bosonic and fermionic observables [49-52]. More recently, it was used to derive a set of equations governing the dynamics of trapped Bose gases [53, 54]. The point is that this principle uses the notion of least biased state, which is the best ansatz compatible with the constraints imposed on the system. For our purposes, we use a Gaussian density operator. The main difference between our approach and the earlier variational treatments is that in our variational theory we not minimize only the expectation values of a single operator such as the free energy in the variational HF and HFB approximation or the thermodynamics potential as is done in variational approach of Bijlsma and Stoof [24]. But our variational theory based on the minimization of an action also with Gaussian variational ansatz. The action to minimize involves two types of variational objects one related to the observables of interest and the other akin to a density matrix [45, 52, 53]. This leads to a set of coupled time-dependent mean field equations for the condensate, the noncondensate and the anomalous average. We call this approach "Time Dependent Hartree-Fock-Bogoliubov" (TDHFB). We have to mention at this point that the TDHFB equations that we derive in this paper are quite general and fully consistent as they do not require any simplifying assumption on the thermal cloud or the anomalous density. They may provide in this sense a kind of generalization to the previously discussed approximations. Moreover, what is important in the TDHFB approach for Bose systems is that there has been no assumptions on weak interactions. Therefore the theory is valid even for strong interactions.

# On the anomalous density for Bose gases at finite temperature

In fact, in many approximations, the anomalous density is always neglected. This is based in the general claim that its contributions are negligibly small. However, Yukalov [42] pointed out that even if it is the case, it is an essential ingredient for the consistency of the formalism. We will go further by using our TDHFB formalism to show that the contribution of the anomalous density is not as small as claimed. On the other hand, the anomalous density profiles seem to have no structure at the centre of the trap for weak interactions. This is in contradiction with what was found in the literature [36, 43] where the HFB-BdG approximation and the classical-field trajectories of the Projected Gross-Pitaevskii equation were used, and where these densities are found to have a "dip".

The paper is organized as follows. In section 2, we review the main steps used to derive the TDHFB equations from the Balian-Vénéroni variational principle. In section 3, the TDHFB equations are applied to a trapped Bose gas to derive a coupled dynamics of the condensate, the non condensate and the anomalous densities. We then restrict to the local densities and discuss the properties of the underlying equations, their relevance to the finite temperature case as well as their relations with other known approximations such as the mean-field Bogoliubov-de Gennes equations. In section 4, we study the behavior of the TDFHB equations in homogeneous Bose gases; in particular we discuss some properties of the normal and anomalous densities as functions of the temperature. Section 5 is devoted to present the static equations and the physical boundary conditions relevant to the trapped Bose gas. We analyze the profiles of the anomalous density for different temperatures both for large and small values of the interaction parameter. We confront our results with the HFB-BdG predictions. Furthermore, we focus on the comparison between the normal and anomalous fractions at low, intermediate and high temperatures. Our conclusions are drawn in section 6.

2. **The variational TDHFB equations**

The time-dependent variational principle of Balian and Vénéroni requires first the choice of a trial density operator. In our case, we will consider a Gaussian time-dependent density operator. This ansatz which belongs to the class of the generalized coherent states allows us to perform the calculations since there exists a Wick's theorem, while retaining some fundamental aspects such as the pairing between atoms.

## On the anomalous density for Bose gases at finite temperature

The Gaussian density operator $D(t)$ is completely characterized by the partition function $Z(t) = \text{Tr} D(t)$, the one-boson field expectation value $<\Psi>(\vec{r},t) \equiv \Phi(\vec{r},t) = \text{Tr}\, \Psi(\vec{r})D(t)/Z(t)$ and the single-particle density matrix $\rho(\vec{r},\vec{r}',t)$ defined as

$$\rho(\vec{r},\vec{r}',t) = \begin{pmatrix} <\overline{\Psi}^+(\vec{r}')\overline{\Psi}(\vec{r})> & -<\overline{\Psi}(\vec{r}')\overline{\Psi}(\vec{r})> \\ <\overline{\Psi}^+(\vec{r}')\overline{\Psi}^+(\vec{r})> & -<\overline{\Psi}(\vec{r})\overline{\Psi}^+(\vec{r}')> \end{pmatrix}. \quad (2.1)$$

In the preceding definitions, $\Psi(\vec{r})$ and $\Psi^+(\vec{r})$ are the boson destruction and creation field operators (in the Schrödinger representation) satisfying the usual canonical commutation rules

$$[\Psi(\vec{r}),\Psi^+(\vec{r}')] = \delta(\vec{r} - \vec{r}'), \quad (2.2)$$

and $\overline{\Psi}(\vec{r}) = \Psi(\vec{r}) - <\Psi(\vec{r})>$.

Upon introducing these variational parameters into the BV principle, one obtains dynamical equations for the expectation values of the one and two boson field operators

$$i\frac{\partial <\Psi>}{\partial t}(\vec{r},t) = \frac{\partial E}{\partial <\Psi^+>(\vec{r},t)},$$

$$i\frac{\partial <\Psi^+>}{\partial t}(\vec{r},t) = -\frac{\partial E}{\partial <\Psi>(\vec{r},t)}, \quad (2.3)$$

$$i\frac{\partial \rho}{\partial t}(\vec{r},\vec{r}',t) = -2\left[\rho, \frac{\partial E}{\partial \rho}\right](\vec{r},\vec{r}',t),$$

where $E = \langle H \rangle$ is the mean-field energy. We may notice at this point that the system (2.3) is closed and does not require any further ingredients. The truncation of the full hierarchy is no longer brutally performed but rather obtained by softly restricting the full Hilbert space to the single particle one.

One of the most noticeable properties of the TDHFB equations (2.3) is the unitary evolution of the single particle density matrix $\rho$, which means that the eigenvalues of $\rho$ are conserved. This implies in particular the conservation of the von-Neumann entropy $S = -\text{Tr}\, D \log D$ and the fact that an initially pure state, satisfying $\rho(\rho+1) = 0$, remains pure during the evolution. This property also leads to the conservation of the Heisenberg parameter [52].

$$I(\vec{r},\vec{r}') = \int d\vec{r}'' \left[ \langle \Psi^+(\vec{r})\Psi(\vec{r}'') \rangle \langle \Psi(\vec{r}'')\Psi^+(\vec{r}') \rangle - \langle \Psi^+(\vec{r})\Psi^+(\vec{r}'') \rangle \langle \Psi(\vec{r}'')\Psi(\vec{r}') \rangle \right]. \quad (2.4)$$

## 3. Application of the TDHFB formalism to trapped Bose gases

Let us apply the previous equations (2.3) to a system of trapped bosons interacting via a two-body potential. The grand canonical Hamiltonian may be written in the form

$$H = \int d\vec{r}\, \Psi^+(\vec{r}) \left[ -\frac{\hbar^2}{2M}\Delta + V_{ext}(\vec{r}) - \mu \right] \Psi(\vec{r}) + \frac{1}{2} \int d\vec{r}\, d\vec{r}'\, \Psi^+(\vec{r})\Psi^+(\vec{r}')V(\vec{r},\vec{r}')\Psi(\vec{r}')\Psi(\vec{r}), \quad (3.1)$$

where $V(\vec{r},\vec{r}')$ is the interaction potential, $V_{ext}(\vec{r})$ the external confining field and $\mu$ the chemical potential. For the sake of clarity, we will omit to write explicitly the time dependence whenever evident. Next, we introduce the order parameter $\Phi(\vec{r}) = \langle \Psi(\vec{r}) \rangle$ and the non-local densities

$$\begin{aligned}
\tilde{n}(\vec{r},\vec{r}') &\equiv \tilde{n}^*(\vec{r},\vec{r}') = \langle \Psi^+(\vec{r})\Psi(\vec{r}') \rangle - \Phi^*(\vec{r})\Phi(\vec{r}'), \\
\tilde{m}(\vec{r},\vec{r}') &\equiv \tilde{m}(\vec{r}',\vec{r}) = \langle \Psi(\vec{r})\Psi(\vec{r}') \rangle - \Phi(\vec{r})\Phi(\vec{r}').
\end{aligned} \quad (3.2)$$

where we note that $\tilde{n}(\vec{r},\vec{r}) \equiv \tilde{n}(\vec{r})$ and $\tilde{m}(\vec{r},\vec{r}) \equiv \tilde{m}(\vec{r})$ are respectively the non condensate and the anomalous densities. The energy may be readily computed to yield

$$\begin{aligned}
E = &\int d\vec{r}\, h^{sp}(\vec{r})\left[\tilde{n}(\vec{r},\vec{r}) + \Phi(\vec{r})\Phi^*(\vec{r})\right] + \int d\vec{r}\, d\vec{r}'\, V(\vec{r},\vec{r}') |\Phi(\vec{r})|^2 |\Phi(\vec{r}')|^2 \\
&+ \frac{1}{2} \int d\vec{r}\, d\vec{r}'\, V(\vec{r},\vec{r}')\left[\tilde{m}^*(\vec{r},\vec{r}')\tilde{m}(\vec{r},\vec{r}') + \tilde{n}(\vec{r},\vec{r}')\tilde{n}(\vec{r}',\vec{r}) + \tilde{n}(\vec{r},\vec{r})\tilde{n}(\vec{r}',\vec{r}')\right] \\
&+ \frac{1}{2} \int d\vec{r}\, d\vec{r}'\, V(\vec{r},\vec{r}')\left[\tilde{n}(\vec{r},\vec{r}')\Phi(\vec{r})\Phi^*(\vec{r}') + \tilde{n}(\vec{r}',\vec{r})\Phi^*(\vec{r})\Phi(\vec{r}') + \tilde{n}(\vec{r},\vec{r})\Phi(\vec{r}')\Phi^*(\vec{r}') + \tilde{n}(\vec{r}',\vec{r}')\Phi(\vec{r})\Phi^*(\vec{r})\right] \\
&+ \frac{1}{2} \int d\vec{r}\, d\vec{r}'\, V(\vec{r},\vec{r}')\left[\tilde{m}^*(\vec{r},\vec{r}')\Phi(\vec{r})\Phi(\vec{r}') + \tilde{m}(\vec{r},\vec{r}')\Phi^*(\vec{r})\Phi^*(\vec{r}')\right]
\end{aligned} \quad (3.3)$$

In the equation (3.3), $h^{sp} = -\dfrac{\hbar^2}{2M}\Delta + V_{ext}(\vec{r}) - \mu$ is the single particle Hamiltonian.

Now, one inserts the expression (3.3) in the general equations of motion (2.3) to get the explicit form of the TDHFB equations for a trapped Bose gas:

$$\begin{aligned}
i\hbar\dot{\Phi}(\vec{r}) = &\, h^{sp}(\vec{r})\Phi(\vec{r}) \\
&+ \int d\vec{r}'\, V(\vec{r},\vec{r}')\left[|\Phi(\vec{r}')|^2 \Phi(\vec{r}) + \Phi^*(\vec{r}')\tilde{m}(\vec{r},\vec{r}') + \Phi(\vec{r}')\tilde{n}(\vec{r},\vec{r}') + \Phi(\vec{r})\tilde{n}(\vec{r}',\vec{r}')\right],
\end{aligned} \quad (3.4a)$$

$$\begin{aligned}
i\hbar\dot{\tilde{n}}(\vec{r},\vec{r}') = &\, \left[h^{sp}(\vec{r}) - h^{sp}(\vec{r}')\right]\tilde{n}(\vec{r},\vec{r}') \\
&+ \int d\vec{r}''\, V(\vec{r}',\vec{r}'')\left[a(\vec{r}'',\vec{r}')\tilde{n}(\vec{r},\vec{r}'') + a(\vec{r}'',\vec{r}'')\tilde{n}(\vec{r},\vec{r}') + b(\vec{r}',\vec{r}'')\tilde{m}(\vec{r}'',\vec{r})\right] \\
&- \int d\vec{r}''\, V(\vec{r},\vec{r}'')\left[a(\vec{r},\vec{r}'')\tilde{n}(\vec{r}'',\vec{r}') + a(\vec{r}'',\vec{r}'')\tilde{n}(\vec{r},\vec{r}') + b(\vec{r},\vec{r}'')\tilde{m}(\vec{r}'',\vec{r}')\right],
\end{aligned} \quad (3.4b)$$

# On the anomalous density for Bose gases at finite temperature

$$i\hbar\dot{\tilde{m}}(\vec{r},\vec{r}') = [h^{sp}(\vec{r}) + h^{sp}(\vec{r}')]\tilde{m}(\vec{r},\vec{r}')$$
$$+ \int d\vec{r}'' V(\vec{r}',\vec{r}'')[a(\vec{r}'',\vec{r}')\tilde{m}(\vec{r},\vec{r}'') + a(\vec{r}'',\vec{r}'')\tilde{m}(\vec{r},\vec{r}') + b(\vec{r}',\vec{r}'')(\tilde{n}^*(\vec{r},\vec{r}'') + \delta(\vec{r} - \vec{r}''))],$$
$$+ \int d\vec{r}'' V(\vec{r}',\vec{r}'')[a(\vec{r}'',\vec{r})\tilde{m}(\vec{r}',\vec{r}'') + a(\vec{r}'',\vec{r}'')\tilde{m}(\vec{r}',\vec{r}) + b(\vec{r},\vec{r}'')\tilde{n}(\vec{r}'',\vec{r}')].$$

(3.4c)

In the Eqs. (3.4), the dots denote time derivatives and we have introduced the quantities

$$a(\vec{r},\vec{r}') = \tilde{n}(\vec{r},\vec{r}') + \Phi^*(\vec{r})\Phi(\vec{r}'),$$
$$b(\vec{r},\vec{r}') = \tilde{m}(\vec{r},\vec{r}') + \Phi(\vec{r})\Phi(\vec{r}').$$

(3.5)

It is worth noticing that similar equations have been derived elsewhere using quite different approaches. For instance, Stoof [24] used a variational plus perturbative effective action, Proukakis [35, 39] a truncation of the Heisenberg equations and Chernyak et al. [55] the generalized coherent state representation. The latter approach yields equations very close to ours, but the authors did not pursue further their analysis. For contact potential $V(\vec{r},\vec{r}') = g\delta(\vec{r} - \vec{r}')$, where g is related to the s-wave scattering length $a$ by $g = 4\pi\hbar^2 a / M$. This leads that the integrations in Eqs (3.4) are removed

$$i\hbar\dot{\Phi}(r) = \left(h^{sp} + gn_c(r) + 2g\tilde{n}(r)\right)\Phi(r) + g\,\tilde{m}(r)\,\Phi^*(r),$$ (3.6a)

The equations of motion for $\tilde{n}(r,r')$ and $\tilde{m}(r,r')$ my be written in the compact form

$$i\hbar\frac{d\rho}{dt} = \Im\rho - \rho\Im^+$$ (3.6b)

Where we have defined the 2x2 matrices

$$\Im(r,r') = \begin{pmatrix} h(r,r') & \Delta(r',r') \\ -\Delta^*(r,r) & -h^*(r,r') \end{pmatrix}, \qquad \rho(r,r') = \begin{pmatrix} \tilde{n}(r,r') & \tilde{m}(r,r') \\ \tilde{m}^*(r,r') & \tilde{n}^*(r,r') + 1 \end{pmatrix}$$

and

$$h(r,r') = h^{sp}(r) + 2ga(r',r')$$
$$\Delta(r,r) = gb(r,r)$$

Eqs. (3.6) constitute the TDHFB equations for the contact interaction potential approximation in real space.

It is well known that in the HFB theory the issues of the ultraviolet divergence of the anomalous density arise from the zero-point occupation of quasiparticle modes [36]. In our case, the diverging term appears in the equation (3.4c) and becomes highly non trivial when considers the contact potential. This is precisely the term which leads to UV-divergences in the anomalous density in the HFB theory [36]. Having identified the origin of this divergence, we can habitually eliminate it from the problem, by regularizing the anomalous average. This is achieved by following the method of refs [56, 57].

The equations (3.6) remain complicated even for contact potential where the noncondensate and anomalous densities are nonlocal functions of two spatial points $\tilde{n}(r,r')$ and $\tilde{m}(r,r')$. To proceed further and to investigate the behavior of the various density profiles both for homogeneous and trapped gases, we consider in this article only the quantities for $r = r'$ since these are the most physically accessible.

### 4. Anomalous density for homogeneous Bose gas

In the uniform case ($V_{\text{ext}}(r) = 0$) and for a thermal distribution at equilibrium, the relation (2.4) may rewrite as

$$I_p = (2\tilde{n}_p + 1)^2 - 4|\tilde{m}_p|^2 = \coth^2(\varepsilon_p / 2kT), \quad (4.1)$$

where $\varepsilon_p$ is the Bogoliubov energy spectrum, defined by the expression

$$\varepsilon_p = \left[(E_p + 2gn - \mu)^2 - g^2(n_c + \tilde{m})^2\right]^{1/2}, \quad (4.2)$$

with $E_p = \dfrac{p^2}{2m}$ is the energy of a free particle.

Note that the expression (4.2) can be derived by computing the RPA modes from (3.6a). It is well known that in order to satisfy the Goldstone or the Hugenholtz–Pines [58] theorem, the spectrum (4.2) should be gapless in the long wavelength limit. This is indeed satisfied provided

$$\mu = g(n + \tilde{n} - \tilde{m}), \quad (4.3)$$

where $n = n_c + \tilde{n}$ is the total density.

Moreover, at zero temperature, the relation (4.1) becomes

$$|\tilde{m}_p|^2 = \tilde{n}_p(\tilde{n}_p + 1). \quad (4.4)$$

The equation (4.4) constitutes an explicit relationship between the normal and the anomalous densities at zero temperature and it indicates that the anomalous density and the thermal cloud density are of the same order of magnitude at low temperatures which leads to the fact that neglecting $\tilde{m}$ while maintaining $\tilde{n}$ is a quite hazardous approximation. Finally, as a technical remark, let us note that the dependence of $\tilde{n}$ and

# On the anomalous density for Bose gases at finite temperature

$\tilde{m}$ at zero temperature allows us to eliminate the non-condensate density from the TDHFB equations, therefore reducing the dimensionality of the problem and simplifying the numerical solution.

A straightforward calculation using equation (4.1) leads to a novel form of the normal and anomalous densities as a function of $\sqrt{I_p}$.

$$\tilde{n}_p = \frac{1}{2}\left[\frac{E_p + g(n_c + \tilde{m})}{\varepsilon_p}\sqrt{I_p} - 1\right]$$

$$\tilde{m}_p = -\frac{g(n_c + \tilde{m})}{2\varepsilon_p}\sqrt{I_p} \qquad (4.5)$$

where $\sqrt{I_p} = \coth(\varepsilon_p / 2kT)$.

It is worth noticing that equations (4.5) together with (3.6.a) constitute the generalized HFB equations at finite temperature. This shows that our formalism can be reproduced easily the full HFB equations both at finite and zero temperatures (see Eqs. (4.6)).

It is convenient now to analyze the asymptotic behavior of equations (4.5) with respect to the temperature.

At low temperature $\tilde{n}_p$ and $\tilde{m}_p$ behave as

$$\tilde{n}_p = \frac{1}{2}\left[\frac{E_p + g(n_c + \tilde{m})}{\varepsilon_p} - 1\right]$$

$$\tilde{m}_p = -\frac{g(n_c + \tilde{m})}{2\varepsilon_p} \qquad (T \to 0). \qquad (4.6)$$

At high temperature we can use the asymptotic form $\sqrt{I_p} \cong \frac{2kT}{\varepsilon_p}$. Then, $\tilde{n}_p$ and $\tilde{m}_p$ take the form

$$\tilde{n}_p = \frac{E_p + g(n_c + \tilde{m})}{\varepsilon_p}\left(\frac{kT}{\varepsilon_p}\right) - \frac{1}{2}$$

$$\tilde{m}_p = -\frac{g(n_c + \tilde{m})}{\varepsilon_p}\left(\frac{kT}{\varepsilon_p}\right) \qquad (T \to T_c). \qquad (4.7)$$

# On the anomalous density for Bose gases at finite temperature

To illustrate comparatively the behavior of the normal and anomalous densities, as functions of temperature for a uniform Bose gas, it is useful to introduce the dimensionless variables $\eta = \dfrac{\varepsilon_p}{g(n_c + \tilde{m})}$ and $\tau = \dfrac{kT}{g(n_c + \tilde{m})}$. The equations (4.5) become

$$\tilde{n}(\eta) = \frac{\sqrt{1+\eta^2}}{2\eta}\sqrt{I(\eta)} - \frac{1}{2},$$
$$\tilde{m}(\eta) = -\frac{1}{2\eta}\sqrt{I(\eta)}$$
(4.8)

where $\sqrt{I(\eta)} = \coth\left(\dfrac{\eta}{2\tau}\right)$

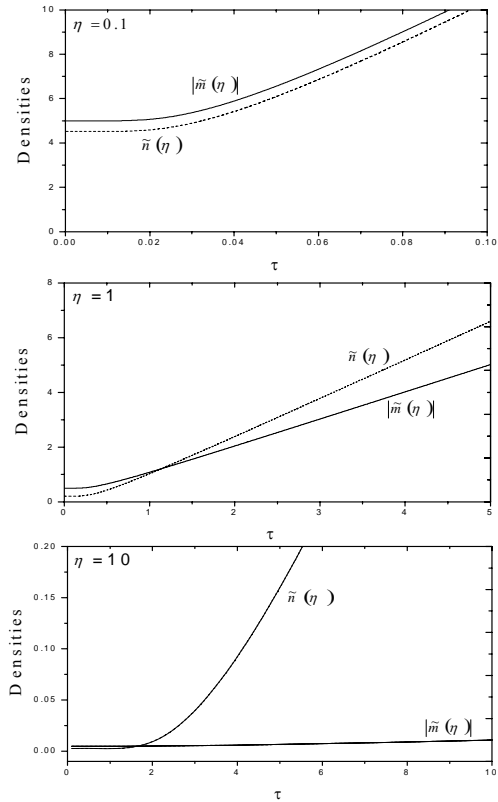

**Figure 1. The absolute value of the anomalous density $\tilde{m}(\eta)$ (Solid line) and the non condensate density $\tilde{n}(\eta)$ (dashed line) vs. the temperature $\tau$ for differents value of $\eta$.**

In this figure we show that the absolute value of the anomalous density is always larger than the noncondensed one for small value of $\eta = 0.1$ at low temperature. Moreover, the absolute value of the anomalous density is comparable with the noncondensed density for $\eta = 1$ ($|\tilde{m}| \approx \tilde{n}$ for $\tau \approx 1.2$). However, at high temperatures, the anomalous density becomes much smaller than the non condensed density. As can be seen from these figures, the behavior of the normal and anomalous densities found here is in good agreement with the recent theoretical results of refs [59-61].

Finally, we can check that omitting the anomalous density, while keeping the normal one, is mathematically inappropriate this is clearly shown in (4.4). From these facts, it is understandable now that neglecting the anomalous density at low temperature for homogeneous Bose gas as one habitually does in the literature is principally unjustified approximation. Let us turn to analyze the situation for trapped Bose gas.

## 5. Anomalous density for a trapped Bose gas

The static TDHFB equations are obtained by setting to zero the right hand sides of Eqs (3.6) in their local form. For numerical purposes, it is convenient to start our treatment with the dimensionless form of the set (3.6). Let us consider a spherical trap with frequency $\omega$, $V_{ext}(r) = \frac{1}{2}m\omega^2 r^2$ and use the harmonic oscillator length $a_{H0} = \sqrt{\hbar/m\omega}$, as well as $a_{H0}^{-3}$ and $\hbar\omega$ as units of length, density and energy respectively. The dimensionless radial distance is $q = r/a_{H0}$. The dimensionless condensed, non-condensed and anomalous densities are respectively $\hat{n}_c = a_{H0}^3 n_c$, $\hat{\tilde{n}} = a_{H0}^3 \tilde{n}$ and $\hat{\tilde{m}} = a_{H0}^3 \tilde{m}$. Therefore, $\hat{n} = \hat{n}_c + \hat{\tilde{n}}$ is the dimensionless total density. The static TDHFB equations can be solved numerically. The numerical method divides into two parts. The first part consists to finding the solutions of the static TDHFB equations that satisfies the boundary conditions summarized as follows: since the full wave function, the normal and the anomalous averages must vanish as $q \to \infty$, the nonlinear term inside the equation (3.6.a) becomes negligible compared to the other terms. Therefore, the two equations of $\hat{\Phi}$ and $\hat{\tilde{m}}$ have the same approximate form $\hat{\Phi} \approx \hat{\tilde{m}} \approx q^{-3/2} e^{\frac{-q^2}{4}}$ for $q \to \infty$. The second part is

# On the anomalous density for Bose gases at finite temperature

used to propagate the solutions of the static TDHFB equations with an adaptable numerical method.

To illustrate our finite temperature formalism, we consider the $^{87}Rb$ gas with the following parameters [30, 62]: $a = 5.82\,10^{-9}$ m, $a_{H0} = 7.62\,10^{-7}$ m and $\hbar\omega = 1.32\,10^{-31}$ J. A convenient dimensionless parameter describing the effective strength of the interactions is $\gamma = Na/a_{H0}$ [20, 63]. We begin by plotting the anomalous density as functions of the radial distance for several values of temperatures for $N = 20000$ atoms i.e. $\gamma = 153$.

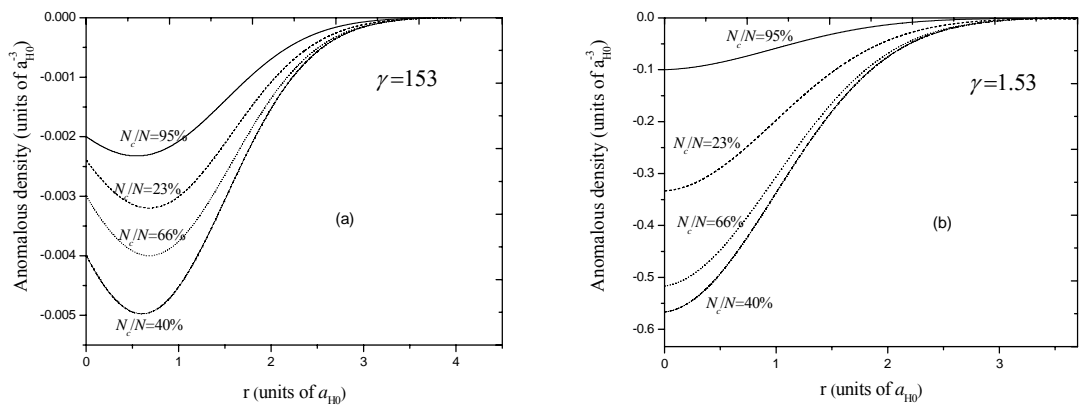

**Figure 2. The anomalous density profiles for various condensate fractions.**

Fig.2 (a) depicts the anomalous density for varying condensation fraction and for $\gamma = 153$. We notice that by decreasing $N_c/N$, $\tilde{m}$ begins to increase in absolute value then decreases when $N_c/N$ approaches 50%. This overall behavior has also been obtained in [36, 43]. The anomalous density remains real and negative value whatever the temperature and the position. The presence of effective repulsive atomic interactions between the atoms, and thus also between the condensate and the thermal cloud, leads to the appearance of a local dip in the anomalous density at the centre of the trap.

In Fig. 2 (b) we show that for a small value of the interaction parameter ($\gamma = 1.53$) the shape of the anomalous density is surprisingly modified in particular we observe that the dip in the neighbourhood of the centre of the trap disappears and the curve takes a Gaussian form. The central anomalous density is lowered for weak interactions. Such a result can be justified by the effect of interactions i.e. for a small number of particles or

# On the anomalous density for Bose gases at finite temperature

small value of $\gamma$, the interactions between the atoms of the condensate and the thermal cloud are lowered this leads automatically that the correlations become weak and that's why the anomalous density remains small and takes a Gaussian shape together with the thermal and the condensate density. We note that the effect of interactions on the anomalous density was studied theoretically earlier by many authors but no one looks how the shape of this quantity varies depending on interactions.

We now turn our attention to analyze more obviously the temperature dependent behavior of the non condensate and the anomalous fractions defined respectively as $\widetilde{N}/N$ and $\left|\widetilde{M}/N\right|$ where $\widetilde{N} = \int d\vec{r}\,\widetilde{n}(r)$ and $\widetilde{M} = \int d\vec{r}\,\widetilde{m}(r)$ are the integrated values of the non condensate and the anomalous densities.

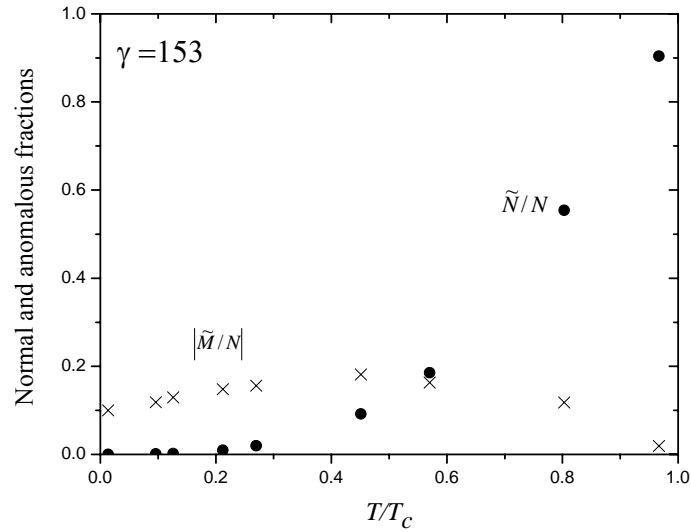

**Figure 3. The normal and anomalous fractions versus reduce temperature for $\gamma = 153$.**

In Fig.3, we plot the non-condensate and the absolute value of anomalous fractions as function of the reduce temperature $T/T_c$ for $\gamma = 153$ (where we follow the method outlined in [36]). It can be seen that $\widetilde{N}/N$ is increasing significantly with increasing temperature and that is why $\widetilde{N}$ reaches to a maximum value close to the Bose-Einstein condensation transition temperature. Furthermore, it is quite interesting to observe that the absolute value of the anomalous fraction $\left|\widetilde{M}/N\right|$ becomes small as the temperature approaches zero or $T_c$. It reaches a maximum at intermediate temperatures ($T \approx 0.5T_c$).

On the other hand, a careful observation into the figure 3 shows that $\left|\widetilde{M}/N\right|$ is larger than $\widetilde{N}/N$ at low temperature ($T \leq 0.5T_c$) which is justified mathematically by our relation (4.4). This result is in disagreement with what is given in literature [36]. In fact, the authors of this reference still based on the idea that the anomalous fraction is always smaller than the normal one for any range of temperature even at low temperatures without any physical or mathematical justifications. Secondly, when the temperature approaches to its critical value the fraction of noncondensed particles becomes close to the unity. That's why at $T \cong T_c$ the anomalous fraction simultaneously with the condensate one become much smaller than the normal fraction. Such feature is well understand and explained in the literature [36, 42, 43, 61]. Although the absolute value of the anomalous and normal fractions are of the same order at intermediate temperatures. It is worth noticing that this behavior persists even for $\gamma = 1.53$ with a small change in the amplitudes of these fractions. Hence, we may infer from these results that the anomalous fraction plays a central role at low temperatures. It is therefore highly unlikely to neglect it for $T < T_c$.

It is well known that the anomalous density is proportional to the s-wave scattering length i.e. the interactions between atoms. In fact, $\widetilde{m}(\vec{r},\vec{r}') = <\overline{\Psi(\vec{r})\Psi(\vec{r}')}> = 0$ for the ideal Bose gas. Indeed, the anomalous average appears in all calculations for Bose system with broken gauge symmetry. Its importance will be highlighted if we compare the expressions, derived with and without taking into account the anomalous averages, with measured quantities, thus measuring the contributions from the anomalous averages. In some cases, the difference is drastic. For instance, the BEC transition is of first order when the anomalous average is not taken into account, while it is second order when we take it into account. Another example is the fact that the compressibility becomes infinite, implying that the system is unstable if the anomalous averages are absent. Moreover, the superfluid transition does not occur, if the anomalous averages are omitted [61].

# On the anomalous density for Bose gases at finite temperature

Since the presence of anomalous average means that particles in the system are correlated, it is important for the current experiments with ultracold atoms to find new methods of characterising the correlated many-body states, such as system near Feshbach resonance, rotating condensates, atoms in optical lattices and low-dimensional systems [64]. An interesting question to ask is whether the anomalous averages can be measured experimentally? Polkovnikov et al in their recent papers [65, 66] argued that this can be done using interference experiments with two independent condensates for any geometry. Nevertheless this technique is not applicable for a single condensate as in our case where there are correlations between the atoms of the condensate and the thermal cloud. Until nowadays, as far as we know, there is no experimental technique which allows for a direct measurement of the anomalous density itself. Experimental data about the anomalous density remain a great challenge for the experimentalists.

## 6. Conclusions

By using a Gaussian density operator, we derive from the time-dependent variational principle of Balian-Vénéroni a set of coupled equations of motion for a self-interacting trapped Bose gas. These equations govern in a self-consistent way the dynamics of the condensate, the thermal cloud and the anomalous average. Our time dependent Hartree-Fock-Bogoliubov (TDHFB) equations generalize in a natural way many of the famous approximations found in the literature such as the Bogoliubov, the Gross-Pitaevskii [67, 68], the Popov [69], the Beliaev [70], the Bogoliubov-de Gennes [36] ZNG equations [22] and others.

The comparison between the normal and anomalous densities at different temperatures is analyzed for homogeneous Bose gas. This analysis shows the significance of the anomalous density compared to the normal one at low temperature [42].

In order to apprehend better the advantages of our approach and owing to its importance to account for many-body effects, we have analyzed the behavior of the anomalous density for trapped Bose gases. We solve numerically the static TDHFB equations in the local limit and for a contact potential. The outcomes of our numerical explorations are numerous. First of all, the numerical resolution of our equations is relatively easy and is not as time-consuming as the HFB-BdG calculations especially for large atom numbers.

# On the anomalous density for Bose gases at finite temperature

For instance, the latter cannot be handled correctly as soon as $N \sim 10^4 - 10^5$. By contrast, there are no such limitations in our case. We recover a well-known theoretical prediction of HFB-BdG [36] since $\widetilde{m}$ increases with the temperature and then decreases as one approaches the transition. Furthermore, we found that the dip of the anomalous density is destroyed for sufficiently weak interactions. Moreover, we show that at low temperatures, the anomalous fraction is larger than the non condensate one. The former necessarily plays a major role in the Bose-Einstein condensation phenomenon. Any approach neglecting the anomalous fraction at low temperatures will inevitably lead to inconsistencies.

The dynamics of the anomalous density will be the goal of our next work with the aim to understand how this quantity evolves in time. Furthermore, to examine more carefully our TDHFB formalism we will focus in the future on the behavior of the anomalous density in dipolar Bose-Einstein condensates at finite temperature both in the static and the dynamic cases.

**Acknowledgments**

We would like to thank J. Dalibard and A. Polkovnikov for many useful comments about this work. Special thanks to G. Shlyapnikov, V.I. Yukalov, J. Walraven, R.Balian and M. Vénéroni.

**References**

[1] N. Bogoliubov, J. Phys. USSR, **11,** 23 (1947)

[2] T.D. Lee and C.N. Yang, Phys. Rev., **112**, 1419 (1958)

[3] P.R. de Gennes, Superconductivity of Metals and Alloys (Benjamin, New York,1966)

[4] R.J. Dodd, M. Edwards, C.J. Williams, C.W. Clark, M.J. Holland, P.A. Ruprecht, and K. Burnett. Phys. Rev.A **54**, 661(1996)

[5] S. Stringari, Phys. Rev A **58**, 2385(1998)

[6] N.P. Proukakis and K. Burnett, Phil. Trans. R. Soc. A **355**, 2235(1997)

[7] G. Baym and C.J. Pethick, Phys. Rev. Lett. **76**, 6 (1996)

[8] V.V. Goldman, I.F. Silvera, and A. J. Leggett, Phys. Rev. B **24**, 2870 (1981)

[9] T. T. Chou, Chen Ning Yang, and L. H. Yu, Phys. Rev. A**55**, 1179 (1997)

[10] W. C. Wu and A. Griffin, Phys. Rev. A**54**, 4204 (1996)

**On the anomalous density for Bose gases at finite temperature**

[11] S. Giorgini, L. P. Pitaeveskii and S. Stringari J. Low Temp. Phys. **109** 309 (1997)

[12] E. Timmermans, P. Tommasini and K. Huang. Phys. Rev. A**55** 3645 (1997)

[13] P. Schuck, X. Vinas, Phys. Rev. A**61**, 43603 (2000)

[14] Y. Castin and R. Dum Phys. Rev. Lett. **77** 5315 (1996)

[15] D. A. W. Hutchinson and E. Zaremba, Phys. Rev. A**57** 1280 (1998)

[16] R. Walser, J. Williams, J. Cooper, and M. Holland, Phys. Rev. A 59, 3878 (1999); R. Walser, J. Cooper and M. Holland Phys. Rev. A**63 ,**013607 (2000)

[17] R. Batch, M. Brewezky and K. Rzazewski. J. Phys. B:Mol. Opt. Phys.**34** 3575 (2001)

[18] S. G. Bhongale, R. Walser and M. J Holland. Phys. Rev. A**66,** 043618 (2002)

[19] B. Jackson and E. Zaremba Phys. Rev. A**66,** 033606 (2002)

[20] S.Giorgini, Phys. Rev. A**57,** 2949 (1998); F. Dalfovo, S. Giorgini, L. P. Pitaevskii and S. Stringari, Rev. Mod. Phys*.* **71**, 463 (1999)

[21] D. A. W. Hutchinson, E. Zaremba and A. Griffin, Phys. Rev. Lett. **78** 1842 (1997)

[22] E. Zaremba, A. Griffin and T. Nikuni, Phys. Rev. A **57**, 4695 (1998); E. Zaremba, T. Nikuni and A. Griffin, 1999, J. Low Temp. Phys. **116**, 277 (1999)

[23] N. P. Proukakis, J. Res. Nat. Inst. Stand. Tech. **101**, 4, 457 (1996)

[24] M. Bijlsma, H. T. C. Stoof, Phys. Rev. A **54**, 5085 (1996); M. Bijlsma, H. T. C. Stoof, Phys. Rev. A**55**, 498 (1997)

[25] C. W. Gardiner and P. Zoller, Phys. Rev. A **61** 033601 (2000)

[26] A.J. Leggett, Rev.:od.Phys **73**, 307 (2001)

[27] A.S.Parkins and D.F.Walls, Phys.Rep.**303**, 1 (1998)

[28] K. Bongs and K. Sengstock, Rep. Prog.Phys.**67**, 907 (2004)

[29] J.O. Andersen, Rev. Mod. Phys. **76**, 599 (2004)

[30] M. H. Anderson, J. R. Ensher, M. R. Mattews C. E. Wieman, E. A. Cornell, Science **269**, 198 (1995)

[31] K.B. Davis, M.O. Mewes, M. R. Andrews, N.J. vanDruten, D.S. Durfee, D.M. Kurn, W. Ketterle, Phys. Rev. Lett. **75**, 3969 (1995)

[32] F. Gerbier, J. H. Thywissen, S. Richard, M. Hugbart, P. Bouyer and A. Aspect, Phys. Rev. Lett. **92**, 030405, (2004)

[33] A. Caracanhas, J.A. Seman, E.R.F. Ramos, E.A.L. Henn, K.M.F Magalhas, K. Helmerson and V.S. Bagnato. J. Phys. B:Mol. Opt. Phys.**42** 145304 (2009)

[34] M.Zawada, R. Abdoul, J. Chwedenczuk, R Gartman, J Szczepkowski, L. tracewski, M Witkowski and W Gawlik. J. Phys. B:Mol. Opt. Phys.**41** 241001 (2008)

[35] N. P. Proukakis, B. Jackson. J. Phys. B: At. Mol. Opt. Phys. **41**, 203002 (2008)

[36] D. A. W. Hutchinson, R.J. Dodd, K. Burnett, S.A. Morgan, M. Rush, E, Zaremba, N.P. Proukakis, M. Edwards, C.W Clark, J. Phys. B **33**, 3825 (2000)

[37] S. Giorgini, Phys. Rev. A**57**, 2949 (1998)

[38] P.O.Fedichev and G.V. Shlyapnikov, Phys. Rev. A **58**, 3146 (1998)

[39] Proukakis, J. Phys. B: At. Mol. Opt. Phys. **34**, 4737 (2001)

[40] S. P. Cockburn, A. Negretti, N. P. Proukakis, and C. Henkel, Phys. Rev. A **83**, 043619 (2011)

[41] A. Griffin, Phys. Rev. B**53**, 9341, (1996); A. Griffin and H. Shi, Phys. Rep. **304** (1998)

[42] V.I. Yukalov, Phys. Rev. E **72**,066119 (2005); V.I. Yukalov and E. P. Yukalova, Laser Phys. Lett. **2**, 505 (2005)

[43] T. M. Wright, P. B. Blakie, R. J. Ballagh, Phys. Rev. A **82**, 013621 (2010)

[44] T. M. Wright, N. P. Proukakis, M. J. Davis, Phys. Rev. A 84, 023608 (2011)

[45] R. Balian and M. Vénéroni, Ann. of Phys. **187**, 29, (1988)

[46] R. Balian, P. Bonche, H. Flocard, and M. Vénéroni, Nucl. Phys. A **428**, 79 (1984); P. Bonche, and H. Flocard Nucl. Phys. A **437**, 189 (1985); J.B. Marston and S.E. Koonin, Phys.Rev.Lett. **54**, 1139 (1985)

[47] O.Eboli, R. Jackiw, and S.-Y.Pi, Phys.Rev.D**37**,3557 (1988)

[48] H. Flocard, Ann. Phys.(NY) **191**, 382 (1989)

[49] R. Balian and M. Vénéroni, Nucl. Phys. B **408**, 445 (1993)

[50] C. Martin, Phys.Rev.D**52**, 7121 (1995)

[51] M. Benarous Ann. Phys. (N.Y) **269**, 107, (1998)

[52] C. Martin, Ann. Phys. (N.Y) **271**, 294 (1999)

[53] M. Benarous and H. Flocard, Ann. of Phys. **273**, 242, (1999)

# On the anomalous density for Bose gases at finite temperature

[54] M. Benarous, Ann. of Phys. **320** 226 (2005)

[55] V. Chernyak, S. Choi, S. Mukamel, Phys. Rev A **67** 053604 (2003)

[56] S. A. Morgan, J. Phys. B **33**, 3847 (2000)

[57] J. O. Andersen, U. Al Khawaja, and H. T. C. Stoof, Phys. Rev. Lett. 88, 070407 (2002)

[58] N. M. Hugenholtz and D. Pines, Phys. Rev. **116**, 489 (1959)

[59] Fred Cooper, Chih-Chun Chien, Bogdan Mihaila, John F. Dawson, and Eddy Timmermans, Phys. Rev. Lett. **105**, 240402 (2010)

[60] A. Rakhimov, E. Ya. Sherman and Chul Koo Kim Phys. Rev. B 81, 020407 (2010)

[61] V.I. Yukalov, Ann. of Phys. **323,** 461, (2008); V.I. Yukalov, Physics Letters A **359**, 712, (2006); V.I. Yukalov, Laser Phys. Lett. **3**, 406, (2006)

[62] D.S. Jin et al, Phys.Rev.Lett.**77**, 992 (1996)

[63] M. Edwards et al, Phys.Rev.Lett.**77**, 1671 (1996)

[64] J. R. Anglin and W. Ketterle, Nature **416**, 211 (2002)

[65] V. Gritsev, E. Demler, and A. Polkovnikov, Phys. Rev. A **78**, 063624 (2008)

[66] A. Polkovnikov, E. Althman, and E. Demler V. Gritsev, Proc.Nat. Acad. Sciences **103**, 6125 (2006)

[67] L.P. Pitaevskii, Sov. Phys. JETP, **13**, 451 (1961)

[68] E.P. Gross, Nuovo Cimento **20**, 454, (1961); J. Math. Phys., 4, 195 (1963)

[69] V. N. Popov, Sov. Phys. JETP **20**, 1185, (1965)

[70] S.T.Beliaev, Sov. Phys. JETP **7** 289, (1958)